# An improved semantic similarity measure for document clustering based on topic maps

Muhammad Rafi[1], Mohammad Shahid Shaikh[2]

[1]Computer Science Department, NU-FAST, Karachi Campus
Pakistan
[1]muhammad.rafi@nu.edu.pk

[2]Electrical Engineering Department, NU-FAST, Karachi Campus
Pakistan
[2]shahid.shaikh@nu.edu.pk

*Abstract*— **A major computational burden, while performing document clustering, is the calculation of similarity measure between a pair of documents. Similarity measure is a function that assigns a real number between 0 and 1 to a pair of documents, depending upon the degree of similarity between them. A value of zero means that the documents are completely dissimilar whereas a value of one indicates that the documents are practically identical. Traditionally, vector-based models have been used for computing the document similarity. The vector-based models represent several features present in documents. These approaches to similarity measures, in general, cannot account for the semantics of the document. Documents written in human languages contain contexts and the words used to describe these contexts are generally semantically related. Motivated by this fact, many researchers have proposed semantic-based similarity measures by utilizing text annotation through external thesauruses like WordNet (a lexical database). In this paper, we define a semantic similarity measure based on documents represented in topic maps. Topic maps are rapidly becoming an industrial standard for knowledge representation with a focus for later search and extraction. The documents are transformed into a topic map based coded knowledge and the similarity between a pair of documents is represented as a correlation between the common patterns (sub-trees). The experimental studies on the text mining datasets reveal that this new similarity measure is more effective as compared to commonly used similarity measures in text clustering.**

## I. Introduction

A similarity measure is a function that assigns a numerical value between 0 and 1 to a pair of objects. The value zero represents that the two objects are totally different, while a value of one suggests that the two objects are identical under the feature sets that are used to represent similarity functions. Measuring the similarity between a pair of documents is becoming increasingly important task as the document collections are growing rapidly. Whether one is clustering a corpus of documents, or searching for relevant documents related to a query text /example document, similarity measure is critical and unavoidable for performing these tasks. Accurate clustering is mainly based on the effectiveness of the defined similarity measure. The function for similarity measure should be easy to compute, it should implicitly captures the relatedness of the documents, and it should also be explainable.

There are several similarity measures suggested by different researchers for the task of document clustering. The most basic one is Euclidean measure that uses the distance metric to compute the similarity of the documents. Another widely used similarity measure is cosine similarity measure, it uses the documents that are represented in vector space and it calculates the angle between the directional vectors of these documents. These two measures do not consider the semantic of the terms (words) in the document. In order to calculate the semantic similarity, researchers suggested WordNet based semantic similarity; they first extract the semantic words from documents and build the semantic class hierarchy of the terms to calculate a similarity based on shared concepts hierarchy.

In this paper we propose a novel similarity measure based on topic maps representation of documents. A topic map is a technology that is becoming a standard for encoding knowledge and connecting the relevant knowledge. We transformed the documents into topic maps based data structures, which capture the semantic of the documents inherently. We also present experimental results to demonstrate that the proposed measure is superior to previously suggested, and frequently used similarity measures for the task of document clustering. The experimental results obtained for clustering clearly indicate that the topic maps based similarity measure is significantly better than existing methods for computing document similarity.

## II. State of the Art

The very first work that studied the impact of similarity measure in the task of clustering was performed by Strehl et al.[1]. They used the YAHOO datasets which are already categorized by human experts in different categories. In order to compare different similarity measures they have performed several different clustering algorithms using these measures. They have used four generally used similarity measures: (i) Euclidean, (ii) Cosine, (iii) Pearson correlation and (iv) Extended Jaccard. The clustering algorithms that they employed were (i) Generalized K-mean, (ii) self-organizing features map, (iii) hyper-graph partitioning and (iv) weighted-graph partitioning. They employed statistical test to ensure the significance of the results of their experiments. The outcome of the experiments established the fact that Extended Jaccard and Cosine similarity are very close to human performed results on categorizations. The work performed in this study is comparable to the work of Strehl et al. In this experiment, a newly proposed similarity measure based on topic maps representation of the documents, along with several of best performed similarity measures of document clustering are compared in the same fashion. This study also extends their work on study the impact of similarity measures to clustering of generalized datasets. Another very closely related work is from Anna Huang [2], it also uses the similar experiment to come up in concluding on effective similarity measure. Our work is different in the sense that we worked on Hierarchical clustering rather than partitional. The paper uses averaged Kullback-Leibler Divergence one of the asymmetric measure, which we have not used in our experiment due to its asymmetric nature. The work of Lin [3] also used the KL-Divergence; this measure is frequently used in word sense-disambiguation. It better represents the similarity of word and their collocation for similarity. One further direction to capture the semantic relatedness of the documents in similarity, different authors suggested to enhance or enrich the document representation with external knowledge base like the work in [4][5]. In [4] the authors suggested that the traditional bag-of-word cannot be very effective in representing the relatedness between documents. They have suggested the enrichment process by adding the relevant terms from WordNet, which were very effective in improving the quality of the clustering results. Further to this, Shady at el. [6][7] suggested the semantic-based term analysis. This approach uses the semantic role labeler for each sentence, then a noise filtering process, removes the common and less semantic words. The terms which are extracted as concept terms are used to represent documents in a compact form. Their semantic similarity measure is calculated based on matching concepts and their hierarchy in the two documents. This work is very closely related to ours, as we are also using the semantic terms via an external topic map library Wandora [8]. Topic Maps [9] becoming a standard for describing knowledge structures and using it to support later the find ability of that coded knowledge. The main emphasis of topic maps structures is to develop a vis-à-vis relation among the knowledge contents. The hierarchical structures of topic maps are very identical to concepts-hierarchy in Shady's work.

A topic map is a representation of a set of assertions about one or more subjects. There are in fact three kinds of assertions topic names, occurrences and associations. The information structure of Topic Maps model [10] is used to structured information in topic maps format. The major benefits of using topic maps for document clustering task are (1) it can reduce the size of document (2) it can capture the topic related information from the document in an structured form (3) the inherit nature of arbitrary and robust information merging and (4) it can easily handle the semantic topics and its hidden relationship and associations.

The different is that our similarity measure uses the common sub-structures of the documents topics and its hierarchy to calculate the semantic similarity between the document pairs. We argue that it is more effective as it captures the semantic of the documents as well as the context based on multi-topic association of the terms that are being used in a pair of documents. if the documents are very short an effective approach is suggested by Wen-tau et al in [11]. The final step clustering has been performed by using hierarchical agglomerative clustering [12].

### III. DOCUMENT REPRESENTATION

Document representation is sensitive to the clustering process. A document is parsed to extract sentence and each sentence is process by Wandora [8], which transform each sentence to a topic map. The document is then compactly represented as a collection of topics along with occurrence and association of topics. Wandora is an application for applying data analysis techniques on topic maps and is able to generate topic maps supporting various data representations. It uses an online plug-in, integrated with a service Open Calais which proved to be very useful in generating Topic maps, taking plain text files as input and returning topic maps based on the information present in these text files. The topic maps are then exported into XTM format using the Wandora's export Utility. A compact document now contains different topic maps structures that are expressed in the document. These are the different distinct trees of the topics and their association. Xpath Queries were used to extract relevant topics, tags and their values from the XTM Files.

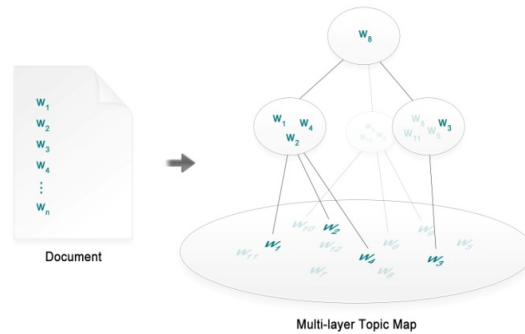

Fig. 1 Example of Multi-Layer Document based on Topic map

### IV. PROPOSED SIMILARITY MEASURE

A Free-Tree based pattern mining and frequency counting algorithm is used to generate the statistical information for the corpus. Let Di and Dj be the two documents from the document corpus D. Again, let the two documents be represented in Topic Maps (as tree) format, as outlined in previous section, as TMdoc1 and TMdoc2. The similarity in trees is complex problem; in order to reduce the complexity of this we introduced the following constraints. Let TMdoc1 contain n nodes and TMdoc2 contain m nodes. Let's number these nodes by following top-to-bottom and left-to-right order (level-breadth- wise), so starting from root which has number 1, the left node of root has number 2 and right node of root has number 3 and so on. Hence the two tree mapping TMdoc1 and TMdoc2 is an ordered pair (i,j), and there will be nxm such pairs. These pair satisfies the following three conditions.

Condition 1: for all pair (i1,j1) and (i2,j2). If i1==i2 and j1==j2 this means that each node is involved in the mapping.

Condition 2: If TMdoc1.Node(i1) is an ancestor of TMdoc1.Node(i2) implies that TMdoc2.Node(j1) is also an ancestor of TMdoc2.Node(j2). It means that the hierarchical order must be preserved for all corresponding nodes.

Condition 3: If TMdoc1.Node(i1) is on the left side of TMdoc1.Node(i2) im-plies that TMdoc2.Node(j1) is also an on the left side of TMdoc2.Node(j2). It means that the sibling order must be preserved for all corresponding nodes.

The proposed similarity measure between these two documents can be defined by the following algorithm:

A. *Algorithm –topic map based similarity for a pair of documents*

Algorithm: Topic Map based similarity measure for a pair of documents
Input: Topic Maps based Tree representation of Documents D1 (V1, E1, root (D1)) and D2 (V2, E2, root (D2))
Output: The nodes Vr of such that there is a root persevering common tree of given pair of documents.
Method:

1. For Vr := 1 to n do
2. S:= { φ}
3. For i= 1 to k do
4. S= {A U B | A ε S, B ε S(Vi)}
5. SΔ= φ;
6. For all v ε V1 such that Label(V1) = Label (V2) do
7. {
8. {If Children(V) ε S then SΔ= SΔ U { v} ;}
9. If (root(D1)) ε SΔ then output V;
10. S(w):= S U SΔ;
11. }

Using the algorithm above a document to document similarity matrix is created. This matrix is finally used to apply Hierarchical Agglomerative clustering.

## V. EXPERIMENTAL STUDY

Evaluating the effectiveness of the proposed similarity measure, we define an experimental setup for performing clustering by using hierarchical agglomerative clustering technique. Clustering is an unsupervised approach which intrinsically based on the sense of similarity which is often subjective in nature. Hence the results produced by clustering algorithms are often compared with manually created category labels using purity of the produced clusters. The motivation behind this evaluation is that in this way we will be able to know, that the unsupervised clustering algorithm is how best in replicating human intelligence. A clustering algorithm is performing excel-lent if the result is consistent with manually created categories. We have selected five different datasets to be compared on the proposed similarity measure. A detail comparison of the proposed similarity measure will be performed with certain well-known and frequently used similarity measure in the literature.

A. *Datasets*

The data for performing this experiment comprises of five datasets, which are selected from standard datasets used in text mining research. These datasets are quite diversified; these are from different sources, from different application, different type of documents, and they contain different number of categories. The characteristics of these five datasets are summarized in the table below:

Table 1: Description of the datasets

| Dataset | #-of-docs | # of classes | Terms | Average-class size |
|---|---|---|---|---|
| NEWS20 | 21000 | 20 | 29550 | 1215 |
| Reuters | 1504 | 13 | 2886 | 131 |
| Webkb | 8230 | 7 | 20650 | 1050 |
| Classic | 7090 | 4 | 12100 | 1774 |
| OSHUMED | 7300 | 13 | 28500 | 187 |

B. *Evaluation*

We justify the effectiveness of our proposed similarity measure by using standard cluster quality measures like Purity and Entropy. The purity of an ideal cluster, cluster which only contains documents from a single category, is equal to 1. We have used this measure to calculate an effective similarity function. An effective similarity must increase the purity of clusters produced based on it. Purity can be thought of as the prediction rate through which a dominated class is learned when it is increasing.

The entropy of an ideal cluster contrary to purity should be close to zero. An effective similarity measure decreases the entropy of the clusters produced based on it. Entropy is more reliable in gauging the effectiveness of similarity measure as it considers the overall distribution of all categories in the clustering results. Further to this, the purity and entropy are independent of the actual results of the clusters. Even a pair of clusters produced by two different similarity measures can have purity measure very close and their entropy can be used to decide the effectiveness of the similarity function, if it has the lowest entropy value.

*1) Purity* : Purity can be defined as the maximal precision value for each class j, We compute the purity for a cluster j as

$$purity(j) = \frac{1}{c_j} max(c_{ij})$$

We then define the purity of the entire clustering result as:

$$Purity = \sum_j \frac{c_j}{N} purity(j)$$

Where $N = \sum_j c_j$ i.e. the sum of the cardinalities of each cluster, Note that we use this quantity rather than the size of the document collection for computing the purity.

2) *Entropy:* Entropy measure how homogenous each cluster j is. It can be calculated by the following formula:

$$Ei = -\sum_{j \in L} presision(i,j) * \log(precision(i,j))$$

The total entropy for a set of cluster is calculated as the sum of entropies for each cluster weighted by the size of each cluster:

$$Entropy_C = \sum_{i \in C}\left(\left(\frac{Ni}{N}\right) * Ei\right)$$

We need to maximize the purity measure and minimize the entropy of clusters in order to accomplish high quality clustering results.

## VI. RESULTS

The experimental results on different similarity measures on the five selected datasets are presented in the Table 2 and Table 3. The Table 2 shows the average purity of the clusters produced by using different similarity measures.

Table2: Purity with different similarity measures

| Dataset | Euclidean | Cosine | Jaccard | KLD | TM-sim |
|---|---|---|---|---|---|
| NEWS20 | 0.23 | 0.54 | 0.5 | 0.38 | **0.56** |
| Reuters | 0.53 | 0.78 | 0.75 | 0.77 | **0.84** |
| Webkb | 0.45 | 0.68 | 0.57 | 0.75 | **0.84** |
| Classic | 0.56 | 0.86 | 0.86 | 0.89 | **0.94** |
| OSHUMED | 0.47 | 0.72 | 0.84 | 0.84 | **0.89** |

The same result in graphical form.

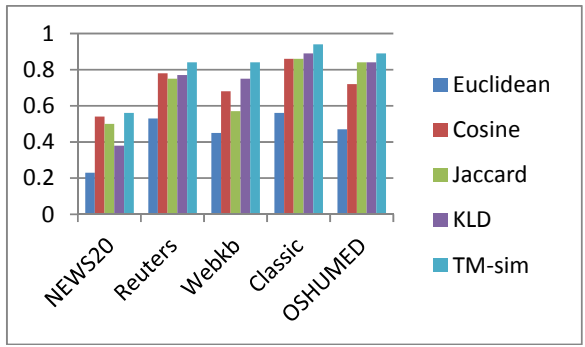

Fig 2: Graphical representation of purity

In Table 3, the average entropy is calculated based on different similarity measures on the selected five datasets.

Table 3: Entropy with different similarity measures

| Dataset | Euclidean | Cosine | Jaccard | KLD | TM-sim |
|---|---|---|---|---|---|
| NEWS20 | 0.95 | 0.51 | 0.52 | 0.54 | **0.35** |
| Reuters | 0.6 | 0.27 | 0.33 | 0.26 | **0.21** |
| Webkb | 0.9 | 0.66 | 0.68 | 0.44 | **0.4** |
| Classic | 0.78 | 0.29 | 0.2 | 0.3 | **0.2** |
| OSHUMED | 0.43 | 0.32 | 0.26 | 0.21 | **0.2** |

The same result in graphical form.

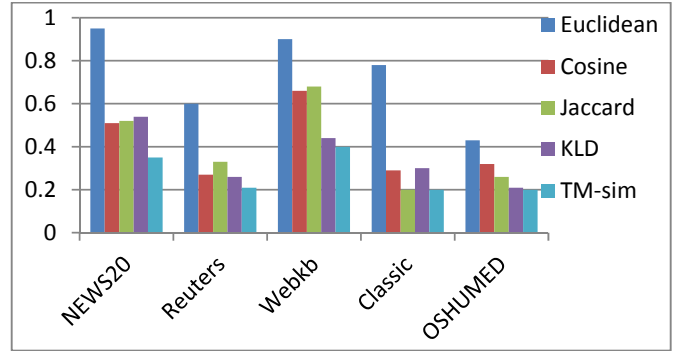

Fig 3: Graphical representation of entropy

## VII. CONCLUSION

In this experimental studies we concludes that the proposed topic map based similarity measure is quite effective in clustering documents collection, as it produced more coherent clustering as compared with human categorized structures. Inspecting the results on purity gives the the fact that KLD similarity performed second best in all datasets, except on NEWS20. This is due to the fact that NEWS20 dataset has very closely related categories and the documents of this collection are also of very short length. The cosine similarity performed fairly well in almost all datasets; traditionally it was a choice to cluster the document collection. The results on entropy also confirm our finding that topic-map is an excellent choice to represent documents for clustering. The entropy result on NEWS20 dataset clearly indicates that this representation is helpful in deciding the overall distribution of the collection and deciding about the actual category of the document. In the future, there can be several directions for research like comparison of documents representation and performance of clustering, different type of clustering algorithms can also be used to investigate the effect of these measures.

ACKNOWLEDGMENT

We would like to thanks anonymous reviewers for their healthy comments and suggestions. We also like to acknowledge the help and support from National University of Computer & Emerging Sciences-(FAST-NU), Karachi campus for carrying out the research on this.

# REFERENCES

[1] Strehl, J. Ghosh, and R. Mooney. Impact of similarity measures on web-page clustering. In AAAI-2000: Workshop on Artificial Intelligence for Web Search, July 2000.

[2] A. Huang, Similarity Measures for Text Document Clustering. In Proceedings of the New Zealand Computer Science Research Student Conference (NZCSRSC'08), Christchurch, New Zealand, 2007

[3] J. Lin. Divergence measures based on the Shannon entropy. IEEE Transaction on Information Theory, 37(1): pp. 145–151, 1991.

[4] T. Pedersen, S. Patwardhan, J. Michelizzi. WordNet::Similarity—measuring the relatedness of concepts. In Proceedings of the 19th National Conference on Artificial Intelligence (AAAI), San Jose, CA, pp. 144–152. 2004.

[5] D. Mccarthy. Relating WordNet senses for word sense disambiguation. In Proceedings of the ACL Workshop on Making Sense of Sense. Trento, Italy. Pp. 17–24. 2006.

[6] S. Shehata: A WordNet-Based Semantic Model for Enhancing Text Clustering. ICDM Workshops 2009: 477-482. 2009.

[7] S. Shehata, F. Karray, M. S. Kamel. An Efficient Model for Enhancing Text Categorization Using Sentence Semantics. Computational Intelligence, 26(3): pp. 215-231, 2010.

[8] Pepper, S., "Topic Maps," Encyclopedia of Library and Information Sciences, Third Edition 2010

[9] Maicher, L.; Garshol, L.M.; Eds. Scaling Topic Maps; In Third International Conference on Topic Maps Research and Applications, TMRA 2007, Leipzig, Germany, October 2007; Springer-Verlag: Berlin, Heidelberg

[10] Document on Wandora Implementation and Usage. Available: http://www.wandora.org/wandora/wiki/index.php?title=Documentation

[11] W.-T. Yih, C. Meek: Improving Similarity Measures for Short Segments of Text. AAAI 2007: pp. 1489-1494, 2007.

[12] B. C. M. Fung, Hierarchical Document Clustering, Master Thesis, Dept. Computer Science, Simon Fraser University, Canada, 2002.